# Investigation of Electron-Phonon Coupling in Epitaxial Silicene by *In-situ* Raman Spectroscopy

Jincheng Zhuang,[1] Xun Xu,[1] Yi Du,[1,*] Kehui Wu,[2] Lan Chen,[2] Weichang Hao,[1,3] Jiaou Wang,[4] Wai Kong Yeoh,[1] Xiaolin Wang,[1] and Shi Xue Dou[1]

[1]*Institute for Superconducting and Electronic Materials (ISEM), University of Wollongong, Wollongong, NSW 2525, Australia*

[2]*Institute of Physics, Chinese Academy of Science, Haidian District, Beijing 100080, P. R. China*

[3]*Center of Materials Physics and Chemistry, and Department of Physics, Beihang University, Beijing 100191, P. R. China*

[4]*Beijing Synchrotron Radiation Facility, Institute of High Energy Physics, Chinese Academy of Sciences, Beijing 100049, P. R. China*

*To whom correspondence should be addressed: yi_du@uow.edu.au

Abstract

In this letter, we report that the special coupling between Dirac fermion and lattice vibrations, in other words, electron-phonon coupling (EPC), in silicene layers on Ag(111) surface was probed by an *in-situ* Raman spectroscopy. We find the EPC is significantly modulated due to tensile strain, which results from the lattice mismatch between silicene and the substrate, and the charge doping from the substrate. The special phonon modes corresponding to two-dimensional electron gas scattering at edge sites in the silicene were identified. Detecting relationship between EPC and Dirac fermion through the Raman scattering will provide a direct route to investigate the exotic property in buckled two-dimensional honeycomb materials.

Silicene, the silicon-based counterpart of graphene, has attracted enormous interest due to its unique characteristics and a wide range of promising applications [1-4]. Theoretical simulations [5,6] predicted, as was very recently verified by experimental work [7], that silicene possesses a graphene-like electronic structure, in which the charge carriers behave as massless Dirac fermions due to the linear electronic dispersion. The strong spin-orbital coupling (SOC) indicates that silicene could show a robust quantum spin Hall (QSH) effect [8]. The interactions between electrons and quantized lattice vibrations, known as electron-phonon coupling (EPC) can facilitate understanding many physical phenomena in silicene including transport behaviour, Kohn anomalies and possible superconductivity. Despite theoretical calculations [9] predicted that free-standing silicene possess unique EPC features due to low-buckled (LB) atomic arrangement, it still needs clarification in experiments. However, it has been an experimental challenge to study and tune EPC in this silicon-based 2D Dirac fermion system because silicene is unstable under ambient condition and *in-situ* investigations are highly desirable. Raman spectroscopy is an insightful tool to probe EPC in a single atomic layer, and the phonon dynamics associating with the 2D Dirac fermions [10]. In silicene, the long wavelength optical $E_{2g}$-phonon mode at $\Gamma$ point of the Brillouin zone (BZ), which corresponds to the relative displacement of non-equivalent neighbour silicon atoms [9,11], is of particular interest. Any perturbations due to this buckled structure will effectively induce the direct electronic transitions across the Dirac point, that is, $E_{2g}$ phonons couple to the low energy excitations.

In this letter, we report on an *in-situ* Raman scattering studies of the phonon modes in epitaxial silicene with different reconstructions on Ag(111) surface at low temperatures. We reveal that the EPC in silicene can be effectively tuned by strain and doping effects, which is demonstrated by shifting of $E_{2g}$ phonon mode. Although depressed by electron doping, EPC in silicene can be significantly enhanced by tensile strain. In addition, our Raman experiments reflect Dirac fermion characteristic in this unique low-buckled 2D material, which is indicated by unique phonon modes attributing to electron scattering at edge sites.

All samples used in this work were fabricated in a preparation chamber supplied with a low-temperature STM/scanning near-field optical microscopy system (LT-STM-SNOM, SNOM1400, Unisoku Co.). Clean Ag(111) substrates were prepared by argon ion sputtering and annealed at 800 K for several cycles. The silicene monolayers were then grown on the Ag(111) surfaces by evaporation of silicon from a heated silicon wafer. *In-situ* Raman spectra were acquired on the silicene samples for the same areas as in the STM measurements. The Raman laser irradiation ($\lambda$ = 532 nm) was delivered through a single-mode optical fibre into the measurement chamber of the STM-SNOM system. All the measurements were carried on in ultrahigh vacuum (UHV) at 77 K. Angle-resolved photoemission spectroscopy (ARPES) characterizations were performed at photoelectron spectroscopy station in Beijing Synchrotron Radiation Facility (BSRF) using a SCIENTA R4000 analyzer. A monochromatized He-I light source (21.2 eV) was used for the band dispersion measurements.

Monolayer silicene with different phases can be obtained by varying the Si coverage and the Ag(111) substrate temperature during deposition. A comprehensive study on the growth mechanism of silicene on the Ag(111) surface has been recently reported [12]. Fig. 1 shows typical STM images of silicene layers in different phases that were grown on Ag(111). Epitaxial silicene exhibits a mixed $\sqrt{13}\times\sqrt{13}$/ 4×4 reconstruction with respect to 1×1 Ag(111) (or 3×3 reconstruction with respect to 1×1 silicene) (lattice constant $a$ = 1.14 nm) in the first monolayers when the substrate temperature was kept at the temperature range from 450 K to 480 K during deposition, as shown in Fig. 1(a). From the second layer, only $\sqrt{3}\times\sqrt{3}$ reconstruction with respect to 1×1 silicene with much smaller lattice constant ($a$ = 0.64 nm) can be observed in top layer of silicene [Fig. 1(b)]. Because $\sqrt{3}\times\sqrt{3}$ multilayer silicene can be fabricated on a 4×4 silicene layer, as demonstrated in previous work [11,13], and on a √13×√13 silicene layer, as shown in this work, suggesting that both 4×4 and √13×√13 silicene can be regarded as buffer silicene layer for the growth of √3×√3 silicene. By carefully maintaining the substrate temperature at 550 K during deposition, √3×√3 silicene can be also epitaxially grown on a bare Ag(111) surface without buffer silicene layer, as shown in Fig. 1(c).

The typical Raman spectrum of the √13×√13/4×4 layer is shown in Fig. 2(a). The first-order asymmetric peak located at 530 cm$^{-1}$ can be interpreted as the zone-centre $E_{2g}$ vibrational mode, which was predicted in previous theoretical studies [14,15]. A shoulder peak from 495 cm$^{-1}$ to 508 cm$^{-1}$ is ascribed to a quantum confinement effect which is a common feature also for microcrystalline silicon [16] and silicon nanowires [17]. The peak at 230 cm$^{-1}$ is assigned to the "$D$" peak, as its intensity is affected by amount of boundary defects in the √13×√13/4×4 silicene layer. Fig. 2(b) displays the Raman spectra for the √3×√3 silicene with different coverage grown on the √13×√13/4×4 layer. The 'SL" stands for the coverage of the top √3×√3 silicene layer. Based on the fact that the √3×√3 silicene heaps up on the √13×√13/4×4 layer following a Stranski-Krastanov (SK) mode [11-13,18,19], 1 SL samples might contain traces of multilayer √3×√3 silicene. The much stronger $E_{2g}$ mode was observed at 530 cm$^{-1}$. Five Raman peaks at lower frequency from 200 cm$^{-1}$ to 500 cm$^{-1}$ (marked as "$D_1$" to "$D_5$") were observed in low-coverage samples. These peaks disappeared when the coverage was more than 1 SL. The same features were also observed in Raman spectra of √3×√3 silicene monolayer grown on Ag(111) without √13×√13/4×4 layer [Fig. 2(c)], which indicates that the low-frequency Raman modes are not induced by interlayer interactions between silicene layers. ARPES measurements on epitaxial √3×√3 silicene indicate that Dirac point locates at ~ 0.33 eV below Fermi level, as shown in Fig. 3(c). This energy shift is not high enough to block the excitation of electrons caused by photons with energy up to 2.3 eV (532 nm). √13×√13/4×4 silicene, however, exhibits a very strong hybridization with the Ag(111) surface, forming a hybridized metallic surface state [20]. These distinct electronic structures of the silicene layers on Ag(111) surfaces indeed reveal the fact that the interactions between Si and Ag are stronger in √13×√13/4×4 silicene, but very weak in √3×√3 silicene. Therefore, the Raman signal is depressed in √13×√13/4×4 silicene. Whether the observed 2D silicon layers are true silicene or just reconstructions of Si(111) surface is still under debate. We carried out Raman measurement on a Si(111) surface as shown in Fig. 1(a). A clear first-order Raman peak ($E_{2g}$ mode) is shown at 520 cm$^{-1}$, which is lower by 10 cm$^{-1}$ than silicene $E_{2g}$ peak (530 cm$^{-1}$). Two broad Raman peaks at 300 cm$^{-1}$ (2TA mode) and 970 cm$^{-1}$ (2TO mode) are assigned to 2TA and 2TO modes in Si(111), which are both absent in silicene Raman spectra. In addition, no Raman signal of Ag(111) can be detected due to Rayleigh scattering. Hence, the $E_{2g}$ and $D$ Raman peaks of silicene layers demonstrate distinct phonon modes that are different to these in Si(111), demonstrating that the epitaxial silicene layers are not reconstructions of Si(111) surface.

Figure 3(a) displays the enlarged view of $E_{2g}$ peaks for √3×√3 silicene with different coverage. The $E_{2g}$ mode frequency (570 cm$^{-1}$) of free-standing (FS) silicene [21] is shown as a reference. The actual positions of $E_{2g}$ peaks can be obtained by the Gaussian-Lorentz peak fitting, as shown in Fig. 3(b). It is obvious that the Raman $E_{2g}$ peak (530 cm$^{-1}$) was softened in epitaxial silicene. As we known, the $E_{2g}$ mode in 2D materials can be modulated by strain [22-27]. The in-plane Si-Si distance $d_{in}$ can be calculated by the distance $a$ between the centres of two neighbouring honeycomb, as shown in Fig. 1(d), according to $d_{in} = a/3 = 2.2$ Å [1,4]. In view of vertical buckling distance $d_B \sim 0.8$ Å, the real Si-Si bond length of the epitaxial √3×√3 silicene is around 2.35 Å. A tensile strain should be present in the √3×√3 silicene layers by considering a smaller Si-Si bond length of 2.24 Å in FS silicene. The frequency shift of the $E_{2g}$ mode in strained silicene can be described as [24,25]:

$$\Delta\omega = \omega_{strain} - \omega_0 = -\frac{nv\omega_0(a_{strain}-a_0)}{a_0} = b\varepsilon \qquad (1)$$

where $\omega_{strain}$ and $\omega_{FS}$ are the frequencies of the $E_{2g}$ mode in strained and FS silicene, respectively, $n$ is the dimensionality of the material, $v$ is the Grüneisen constant, $b$ is the strain-shift coefficient, and $a_{strain}$ and $a_0$ are the *in-plane* lattice parameters of the strained and FS silicene layers, respectively. The *in-plane* strain, $\varepsilon = (a_{strain} - a_0)/a_0 = \Delta a/a_0$, is 0.05 which was obtained from the Si-Si bond lengths for both the strained and FS silicene. The detailed structural parameters are listed in TABLE 1. The frequency of the $E_{2g}$ mode in silicene downshifts to 520 cm$^{-1}$ by taking coefficient $b$ = -832 cm$^{-1}$ [25,28,29], which is smaller than that in our Raman spectra (530 cm$^{-1}$) but comparable to the $E_{2g}$ mode frequency of Si(111). There must be another factor to affect phonon frequency of $E_{2g}$ mode.

In fact, electron or hole doping in silicene can result in a hardening of the mode frequency in silicene [9,11], *e.g.* $E_{2g}$ mode will shift to higher frequency due to charge doping. ARPES verifies that the Dirac point of √3×√3 silicene grown on Ag(111) is located at ~ 0.33 eV below the Fermi surface due to electron doping from the Ag(111) substrate, as shown in Fig. 3(c). $E_{2g}$ vibration at the zone centre (**Γ** point) couples with Dirac fermions at the zone boundary (**K** points), which is allowed by silicene lattice symmetry. The carriers residing in the honeycomb lattice mutually interact with the $E_{2g}$ mode through dynamical perturbations due to creation and annihilation of virtual long-wavelength electron-hole pairs across the gapless Dirac point. The energy range of the virtual electron-hole pairs allowed by the Pauli principle is decided by the position of the Fermi energy ($E_F$). Since the $E_F$ of silicene is lifted by electron doping, $E_{2g}$ mode in silicene is hardened. The $E_{2g}$ frequency upshifting by about 10 cm$^{-1}$ agrees well with previous simulation results [9]. Fig. 3(d) is a schematic of both strain and doping effects on the $E_{2g}$ Raman peak in √3×√3 silicene.

It is noted that EPC strength depends primary on the phonon frequencies in 2D materials [30,31]. In general, EPC can be characterized by a dimensionless parameter $\gamma = N_F V_{ep}$, where $N_F$ is the electron DOS and $V_{ep}$ is the mean electron-phonon coupling potential at Fermi level [31]. The value of $\gamma$ is proportional to $\omega^{-2}$ ($\gamma \sim \omega^{-2}$) [30,31]. This relation originates from the zero-point oscillation amplitude induced by large deformation and the energy reconstruction in the perturbation theory. Both factors correspond exactly to the stain effect presenting in 2D materials. Therefore, the value of $\omega$ reflects the EPC strength. According to $E_{2g}$ peak shift from 570 cm$^{-1}$ (FS) to 520 cm$^{-1}$ (strained silicene), the EPC in silicene can be enhanced up to 20%. The enhancement of EPC is of particular interest because it gives rise to superconductivity in Bardeen-Cooper-Schrieffer (BCS) superconductors. Our results may support recent observations on the existence of a superconducting gap in silicene layers [32].

In graphene, the armchair and zigzag edges induce two Raman-active $D$ and $D'$ modes due to intervalley and intravalley scatterings of quasiparticles in Dirac cones, respectively [23-35]. As a similar Dirac fermion system, silicene should also possess such features in Raman spectroscopy, as shown in Fig. 4(a) inset. In our results, five distinct Raman peaks at low wavenumber range (220 cm$^{-1}$ to 470 cm$^{-1}$) were observed in the samples with coverage less than 1 SL, as shown in Fig. 2 (b). Fitted spectra of these Raman modes are shown in Fig. 4(a) and Fig. S1, where five peaks denoted as "$D_1$" to "$D_5$". The intensities of these peaks can be scaled well with each other and show a strong dependence on the coverage. Therefore, it suggests that the peaks are most likely associated with edges. In STM results, only two edge angles can be observed, which correspond to two armchair/zigzag arrangements that are 150° zigzag-armchair (ZA) and 120° armchair-armchair (AA), as shown in Fig 4(c) and (d). Due to the low buckled structure of silicene, the structure symmetry is further lowered in contrast to planar graphene. For example, two zigzag edge structures could possibly exist in $\sqrt{3}\times\sqrt{3}$ silicene layers (Fig. S3). One only consists of up-buckled atoms and the other one only consists of down-buckled atoms. Consequently, it is expected that more vibrational modes are induced by various edge arrangements in Raman spectroscopy of silicene, which agrees well with $D_1$ - $D_5$ peaks in our results. The edge-induced Raman peaks reflect the unique buckled characteristic in silicene. Tip-enhanced Raman spectroscopy is expected to advance insight into edge effects on phonon modes in this low-buckled 2D material.

In summary, silicene layers with different structures and coverage have been fabricated, and identified by *in-situ* UHV Raman spectroscopy and STM. The intrinsic phonon modes of silicene layers in different structures are revealed. We found that EPC in silicene can be significantly enhanced due to the lattice mismatch between silicene layers and the substrate. The Raman spectroscopy demonstrates the effects of coverage, strain, charge doping and defects on silicene's phonon modes, and allows unambiguous, high-throughput, nondestructive identification of epitaxial silicene.

This work was supported by the Australian Research Council (ARC) (DP140102581, LE100100081 and LE110100099). Y. Du would like to acknowledge support by the University of Wollongong through a University Research Council (URC) Small Grant. This work is partially supported by the MOST of China (Grants No. 2013CBA01600) and the NSF of China (Grants No. 11334011, 11322431, 51272015). The authors thank Dr. H. Pan and Dr. Tania Silver for valuable discussion.

[1] B. J. Feng *et al.*, Nano Lett. **12**, 3057 (2012).

[2] J. F. Gao, and J. J. Zhao, Sci. Rep. **2**, 86 (2012).

[3] C. -L. Lin *et al.*, Phys. Rev. Lett. **110**, 076801 (2013).

[4] P. Vogt *et al.*, Phys. Rev. Lett. **108**, 155501 (2012)

[5] S. Cahangirov *et al.*, Phys. Rev. Lett. **102**, 236804 (2009).

[6] C. -C. Liu, W. X. Feng, and Y. G. Yao, Phys. Rev. Lett. **107**, 076802 (2011).

[7] L. Chen *et al.*, Phys. Rev. Lett. **109**, 056804 (2012).

[8] M. Ezawa, Phys. Rev. B **87**, 155415 (2013).

[9] J. -A. Yan *et al.*, Phys. Rev. B **88**, 121403 (2013).

[10] N. B. Kopnin and E. B. Sonin, Phys. Rev. Lett. **100**, 246808 (2008).

[11] P. De Padova *et al.*, 2D Mater. **1**, 021003 (2014).

[12] X. Xun *et al.*, The IEEE ICONN 2014 Conference proceedings (Submitted).

[13] P. De Padova *et al.*, Appl. Phys. Lett. **102**, 163106 (2013).

[14] A. Gupta *et al.*, Nano. Lett. **6**, 2667 (2006).

[15] R. Arafune *et al.*, Surf. Sci. **608**, 297 (2013).

[16] Z. Iqbal, and S. Veprek, J. Phys. C: Solid State Phys. **15**, 377 (1982).

[17] B. B. Li, D. P. Yu, and S. -L. Zhang, Phys. Rev. B **59**, 1645 (1999).

[18] P. Vogt *et al.*, Appl. Phys. Lett. **104**, 021602 (2014).

[19] E. Salomon, R. El Ajjouri, G. Le Lay, and T. Angot, J. Phys.: Cpndens. Matter **26**, 185003 (2014).

[20] D. Tsoutsou *et al.*, Appl. Phys. Lett. **103**, 231604 (2013).

[21] J. Yan, Y. B. Zhang, P. Kim, and A. Pinczuk, Phys. Rev. Lett. **98**, 166802 (2007).

[22] F. Speck *et al.*, Appl. Phys. Lett. **99**, 122106 (2011).

[23] J. Röhrl *et al.*, Appl. Phys. Lett. **92**, 201918 (2008).

[24] D. J. Lockwood, and J. -M. Baribeau, Phys. Rev. B **45**, 8565 (1992).

[25] R. -P. Wang *et al.*, Phys. Rev. B **61**, 16827 (2000).

[26] E. Cinquata *et al.*, J. Phys. Chem. C **117**,16719 (2013).

[27] E. Scalise *et al.*, Nano. Res. **6**, 19 (2013).

[28] E. Anastassakis, A. Cantarero, and M. Cardona, Phys. Rev. B **41**, 7529 (1990).

[29] S. Nakashima, T. Mitani, M. Ninomiya, and K. Matsumoto, J. Appl. Phys. **99**, 053512 (2006).

[30] W. McMillan, Phys. Rev. 167, 331 (1968).

[31] C. Si, Z. Liu, W. H. Duan, and F. Liu, Phys. Rev. Lett. 111, 196802 (2013).

[32] L. Chen, B. J. Feng, and K. H. Wu, Appl. Phys. Lett. 102, 081602 (2013).

[33] L. G. Cançado *et al.*, Phys. Rev. Lett. **93**, 247401 (2004).

[34] R. Saito, G. Dresselhaus, and M. S. Dresselhaus, Phys. Rev. B **61**, 2981 (2000).

[35] C. Casiraghi *et al.*, Nano. Lett. **9**, 1433 (2009).

**Figures and figure captions**

**Figure 1**

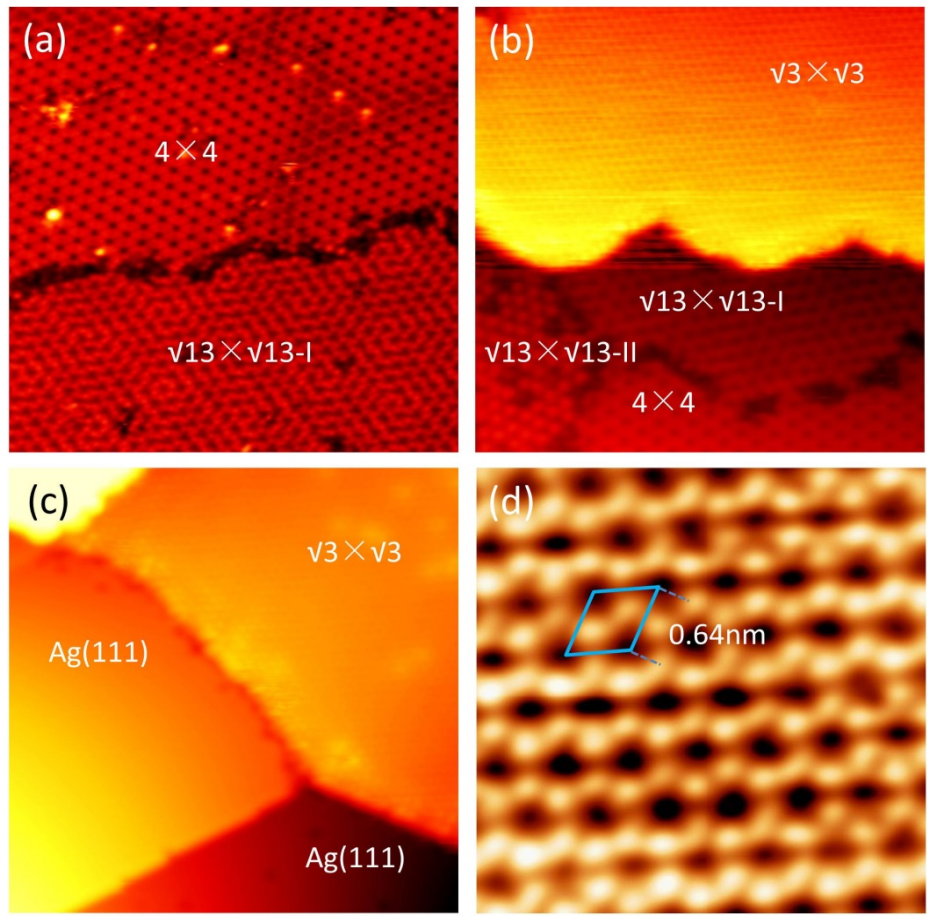

FIG 1. STM images of silicene layers in different phases, (a) mixed √13×√13/4×4, (b) √3×√3 silicene grown on √13×√13/4×4 buffer layer, (c) √3×√3 silicene grown on Ag(111) substrate. (30 nm × 30 nm, $V_{bias}$ = -1.0 V, $I$ = 1 nA) and (d) the enlarged view of √3×√3 silicene. The rhombus stands for the unit cell of √3×√3 silicene, which is used to calculate the lattice parameter of √3×√3 silicene. (5 nm × 5nm, $V_{bias}$ = 0.5 V, $I$ = 4 nA)

**Figure 2**

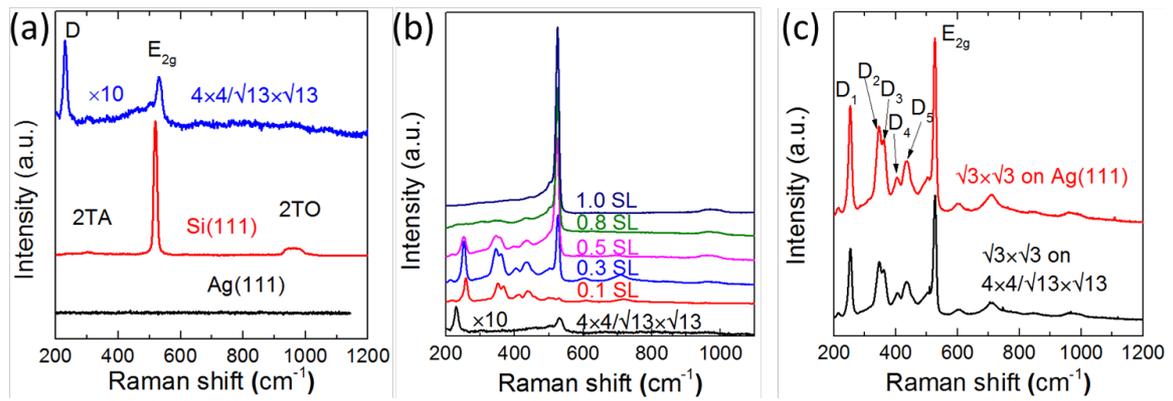

FIG 2(a) Raman spectra of Ag(111) substrate, Si(111), and √13×√13/4×4 buffer layer. (b) Raman spectra of √3×√3 silicene grown on √13×√13/4×4 buffer layer with different coverage. "SL" denotes the coverage of the √3×√3 silicene layer. (c) Raman spectra of √3×√3 silicene layer (coverage of 0.3 ML) grown on Ag(111) surface and √13×√13/4×4 buffer layer, respectively.

**Figure 3**

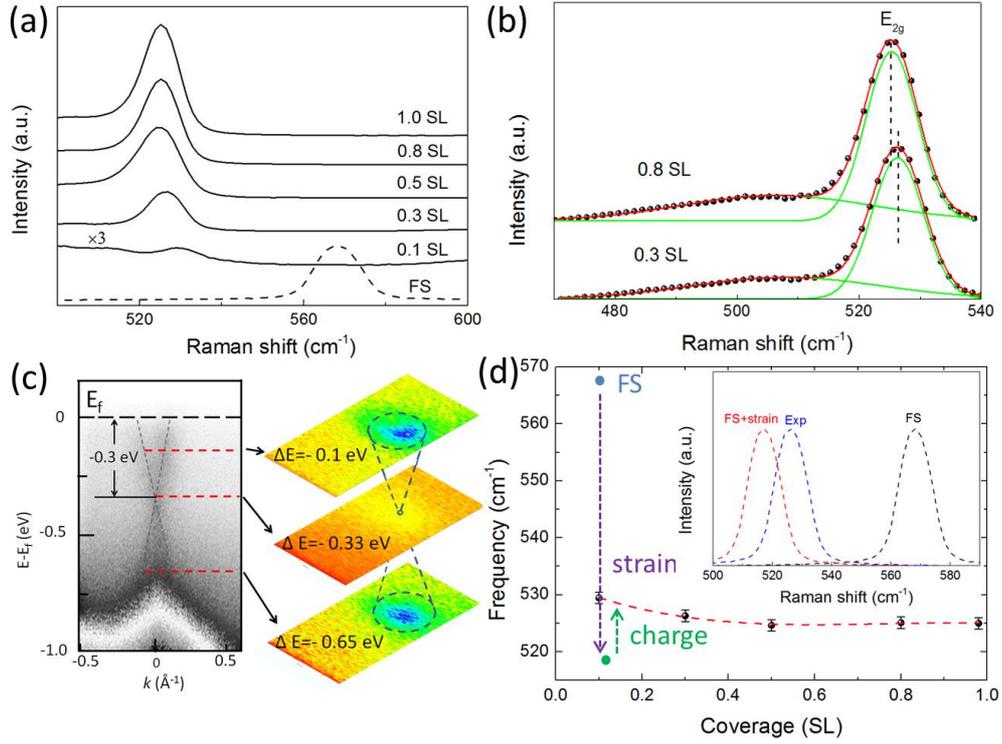

FIG 3(a) The enlarged view ranging from 500 cm$^{-1}$ to 600 cm$^{-1}$ of Raman spectrum of $E_{2g}$ peak for different coverage samples. The $E_{2g}$ mode frequency of FS silicene is plotted as a reference [7]. (b) Fitted results of $E_{2g}$ peak for 0.3 SL sample and 0.8 SL sample. The dashed lines are used to mark the position of $E_{2g}$ mode. (c) ARPES results of epitaxial √3×√3 silicene layer. The left part shows that two faint linear dispersed bands crossed at BZ centre $\Gamma$ point. The right part displays constant-energy cuts of the spectral function at different binding energies verify that both bands origin from a Dirac cone structure, which can be assigned to linear $\pi$ and $\pi^*$ states of √3×√3 silicene. The Dirac point is about 0.33 eV below Fermi level which indicates electron-doping effect attributed to Ag(111) substrate. (d) $E_{2g}$ mode frequency as a function of coverage. Both the (d) and the inset of (d) are sketched to illustrate the strain and doping effects on Raman peak position of $E_{2g}$ mode in √3×√3 silicene, in which the strain effect softens the $E_{2g}$ mode, while charge doping will upshift the $E_{2g}$ mode.

**Figure 4**

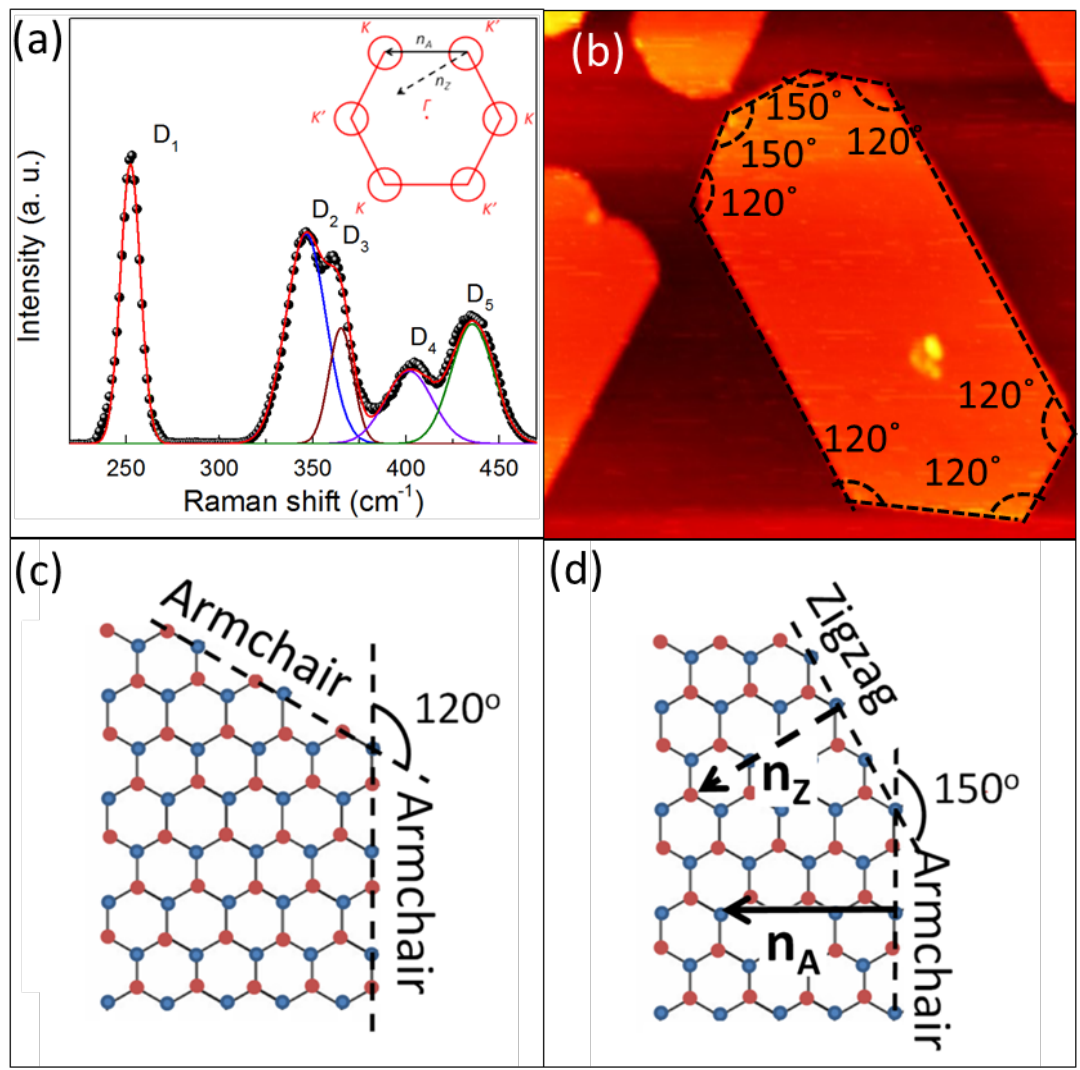

FIG 4(a) Fitted Raman spectra of √3×√3 silicene layers with coverage of 0.3 SL in the frequency range of 220 to 470 cm$^{-1}$, in which Raman peaks due to edges are marked as $D_1$ to $D_5$. (b) STM image of two typical arrangements of edges: armchair/zigzag and armchair/armchair, resulting in two edge angles of 150º and 120º, respectively. (c) and (d) Atomic structures of armchair and zigzag edges. Only the armchair edge supports elastic intervalley scattering of electrons in the Brillouin zone, as indicated in inset of (a).

**TABLE 1.** Detailed structural parameters of free-standing (FS) silicene and low-buckled (LB) silicene: range of Si-Si bond length for experimental results ($d^{exp}_{si-si}$) and calculation results ($d^{cal}_{si-si}$), the average Si-Si length ($d_a$) used for the calculation of *in-plane* strain ($\varepsilon = (d_a(LB) - d_a(FS)/d_a(FS))$) of LB silicene grown on Ag(111) substrate in reward of FS silicene.

| Type | $d^{exp}_{si-si}$ | $d^{cal}_{si-si}$ | $d_a$ | $\varepsilon$ | $E_{2g}$ peak |
|---|---|---|---|---|---|
| FS | N/a | 2.24 Å [9,26] | 2.24 Å | 0 | 570 cm$^{-1}$ [9] |
| LB | 2.32~2.38 Å | 2.28~2.40 Å [2,7] | 2.35 Å | 5 % | 530 cm$^{-1}$ |